\documentclass[showpacs,preprintnumbers,twocolumn,
amsmath,amssymb,groupedaddress,superscriptaddress]{revtex4}
\usepackage{graphicx}
\usepackage{dcolumn}
\usepackage{bm}

\begin{document}


\title{Nuclear skin emergence in Skyrme deformed Hartree-Fock calculations}

\author{P. Sarriguren}
\affiliation{Instituto de Estructura de la Materia, CSIC, Serrano
123, E-28006 Madrid, Spain}

\author{M.K. Gaidarov}
\affiliation{Instituto de Estructura de la Materia, CSIC, Serrano
123, E-28006 Madrid, Spain}
\affiliation{Institute for Nuclear
Research and Nuclear Energy, Bulgarian Academy of Sciences, Sofia
1784, Bulgaria}

\author{E. Moya de Guerra}
\affiliation{Departamento de Fisica Atomica, Molecular y Nuclear,
Facultad de Ciencias Fisicas, Universidad Complutense de Madrid,
E-28040 Madrid, Spain}

\affiliation{Instituto de Estructura de la Materia, CSIC, Serrano
123, E-28006 Madrid, Spain}

\author{A.N. Antonov}
\affiliation{Institute for Nuclear Research and Nuclear Energy,
Bulgarian Academy of Sciences, Sofia 1784, Bulgaria}


\begin{abstract}
A study of the charge and matter densities and the corresponding
rms radii for even-even isotopes of Ni, Kr, and Sn  has been
performed in the framework of deformed self-consistent mean field
Skyrme HF+BCS method. The resulting charge radii and neutron skin
thicknesses of these nuclei are compared with available
experimental data, as well as with other theoretical predictions.
The formation of a neutron skin, which manifests itself in an
excess of neutrons at distances greater than the radius of the
proton distribution, is analyzed in terms of various definitions.
Formation of a proton skin is shown to be unlikely.
The effects of deformation on the neutron skins in even-even
deformed nuclei far from the stability line are discussed.
\end{abstract}


\maketitle

\section[]{INTRODUCTION\label{s:intro}}
The detailed study of the properties of unstable nuclei far from
the stability line has been one of the main goals of nuclear
physics in the last years. Recently, the development of
radioactive ion beam (RIB) facilities in GSI (Germany) and in
RIKEN (Japan) has opened a new field for such study, making possible
the production of a variety of exotic nuclei which may have large
neutron or proton excess.

So far studies have largely dealt with the light nuclei which
became accessible by the use of RIB produced in fragmentation
reactions. Prior experiments have revealed a halo phenomenon (e.g.
in Refs.~\cite{Tani95,Tani85}) whose occurrence is due to both the
small separation energy of the last few nucleons and their
occupation on the orbits with low angular momentum. As well as
halos, heavier systems may lead to formation of pronounced neutron
skins \cite{Suzuki95} due to availability of far more neutrons
than protons in these nuclei.

Immediate determination of the neutron skin thickness usually
involves the precise measurement of the root mean square (rms)
radii of both charge and mass distributions.
Electron-nucleus scattering has proven to be an excellent tool for
the study of nuclear structure. In particular, it has accumulated
much reliable information on charge density distributions of
stable nuclei. Therefore, it is believed that the new facilities
in GSI \cite{GSI2005,Simon2004} and RIKEN \cite{Suda2005} will
provide a good opportunity to study the charge density, and
consequently the proton density distribution, of unstable nuclei
by elastic electron scattering. Unfortunately, a measurement of
the neutron density distributions to a precision and detail
comparable to that of the proton one is hardly possible. The
nuclear matter distribution in $^{6}$He and $^{8}$He has been
determined recently at GSI by using small angle proton scattering
in inverse kinematics at relativistic energy \cite{Alkha97}, and
data has also been collected for $^{11}$Li. It turned out that to
get information on the neutron skin thickness one needs data
obtained with probes having different sensitivities to the proton
and neutron distributions. The methods for extracting
the neutron skin thickness mostly include hadron scattering
\cite{Ray79,Hoff81}, antiprotonic atoms \cite{Agni2001}, parity
violating electron scattering \cite{Donn89,Horow93,Vret2000}, as
well as
giant dipole resonance method \cite{Kras94} and spin-dipole
resonance method \cite{Kras99,Kras2004}.

On the theoretical side, calculations of nuclear charge and matter
radii of exotic nuclei are usually made in the framework of
mean-field approaches, namely Hartree-Fock (HF) method (see for
example \cite{Hofmann98,Richter2003}) or Hartree-Fock-Bogoliubov
(HFB) method including pairing correlations
\cite{Angeli80,Mizu2000,Amos2004,Ant2005,Benn2003,Libert2007}. The
latter predicts well the monotonic increase of the neutron skin
thickness for different chains of isotopes up to the drip line
\cite{Angeli80}. Recently, the self-consistent relativistic
mean-field (RMF) model has been widely applied to both stable
and unstable nuclei (e.g.,
Refs.~\cite{Lala95,Lala99,Wang2004,Wang2006}). Also the
relativistic Hartree-Bogoliubov approach has been employed to
study the nuclear skin thickness in neutron/proton-rich sodium
isotopes \cite{Gamb2001}. Many calculations show that the RMF
model can reproduce with good precision a number of ground-state
nuclear properties including the charge radii \cite{Ren95}. The
charge rms radii were successfully described very recently in
Ref.~\cite{Libert2007}, where a generator coordinate method (GCM)
on top of Gogny HFB calculations was explored.

Theoretical identification of skin and/or halo
structure in neutron-rich weakly bound nuclei, however, is still a
matter of discussion. In Ref.~\cite{Fuku93} a definition
of the neutron skin and its appearance were presented in terms of
spherical HF calculations. The proposed criteria which deal
with proton and neutron densities allowed one to predict
neutron skins in nuclei far from the $\beta$ stability line. It has
been also shown in \cite{Fuku93} that the formation of proton skin
appears to be rather difficult. The Helm model
\cite{Helm56,Sprung92} has been applied in Ref.~\cite{Mizu2000} to
analyze neutron and proton skins, as well as halos, of even-even
Ni, Sn, and Pb isotopes in terms of form factors.
In Ref.  \cite{Mizu2000} three different definitions were proposed for
neutron-proton radii differences. Among them,
the one based on the Helm model was chosen as a measure of the skin.
The latter was shown to have a smooth gradual
dependence on the neutron excess and to be  almost unaffected by shell
effects. The Helm model was used very recently also by Bertulani
\cite{Bertu2007} to investigate electron scattering from light
unstable nuclei.

Hitherto, the different definitions for the skin thickness
mentioned above have been explored within different nuclear
structure models. We would like to emphasize that a comparison of
skins extracted by using various definitions is not very
meaningful unless the same nuclear model is used and this has not
been done in the past. Such an analysis of neutron skins
within a given microscopic nuclear structure model could be very
useful also in respect to demonstrate their expected spreading
when different definitions of the nuclear skin are used.

Another interesting question is to explore how the
neutron skin emerges in the presence of deformation. The
latter is defined by the non-spherical components of the
proton and neutron density distributions. In particular, studies of
deformed exotic nuclei and skins can be found in
Refs.~\cite{hamamoto,stoitsov}.
It is desirable to study the evolution of shape and skin formation,
not only because deformation influences the nuclear rms radii, but
also because of the possible skin anisotropies that may take place.

In the present study, the properties of even-even  Ni
($A$=48--78), Kr ($A$=70--100), and Sn ($A$=100--136) isotopes are
described using the deformed self-consistent mean-field Skyrme
HF+BCS method. We have used three parametrizations of the Skyrme
force, namely SG2, Sk3 and SLy4, which were able to give an
appropriate description of bulk properties of spherical and
deformed nuclei in the past. As in our previous paper
\cite{Ant2005}, we choose some medium and heavy Ni, Kr, and Sn
isotopes because many of these sets, which lie in the nuclear
chart between the proton and neutron drip lines can be formed as
radioactive ions to perform scattering experiments. The main goal
of this study is to clarify theoretically the emergence of the
neutron and proton skins in neutron-rich and neutron-deficient
isotopes, respectively, by testing different definitions for the
skin thickness in the framework of the deformed Skyrme HF+BCS
model. Alternatively to one of the criteria for the neutron skin
proposed in Ref.~\cite{Fuku93} we consider another one which
treats proton and neutron densities in a similar way.
We extend the analysis of nuclear sizes presented in
Ref.~\cite{Ant2005} by performing a more systematic study of a
larger set of exotic nuclei and calculating also neutron skin
thicknesses. The calculated charge rms radii are compared with the
laser or muonic atoms spectroscopy measurements of isotope shifts
performed on Sn \cite{Blanc2002,Blanc2005,Ansel86,Piller90}, Ni
\cite{Pomor95,fricke}, and Kr \cite{Keim95} isotopes. The neutron
skin thicknesses obtained in this paper are compared with the
available experimental data extracted from methods mentioned above
for even-even Sn isotopes with masses from 112 to 124. We also
study whether the emergence of a skin is influenced by the nuclear
shape, an issue that has not been sufficiently studied so far. The
question of skin formation in nuclei having a non-spherical shape
is discussed in detail on the example of Kr isotopes, assuming
axial symmetry.

The paper is organized in the following way. Section II contains
the formalism of the deformed Skyrme HF+BCS method that provides
the model density distributions, form factors, and nuclear radii.
The numerical
results and discussions are presented in Sec. III. Finally, we
draw the main conclusions of this study in Sec. IV.

\section{Deformed Skyrme HF+BCS formalism}

The results discussed in the next sections have been obtained from
self-consistent deformed Hartree-Fock calculations with density
dependent Skyrme interactions \cite{vautherin} and pairing correlations.
Pairing between like nucleons has been included by solving
the BCS equations at each iteration either with a fixed pairing gap parameter
(determined from the odd-even experimental mass differences) or with a
fixed pairing strength parameter.

We consider in this paper the Skyrme force SLy4 \cite{sly4}. We
also show in some instances the results obtained from other
parametrizations, namely Sk3 \cite{sk3} and  SG2 \cite{sg2}
because they are among the most extensively used Skyrme forces and
are  considered as standard references.

Assuming time reversal, the single-particle Hartree-Fock solutions
for axially symmetric deformed nuclei are characterized by the
eigenvalue $\Omega_i$ of the third component of the total angular
momentum on the symmetry axis and by the parity $\pi_i$. The state
$i$ can be written as
\begin{eqnarray}
\Phi_i \left( {\vec R},\sigma ,q\right)& = &\chi_{q_i}(q) \bigg[
\Phi^+_i (r,z) e^{i\Lambda^-\varphi} \chi_+(\sigma) \nonumber \\
&+& \Phi^-_i (r,z) e^{i\Lambda^+\varphi} \chi_-(\sigma) \bigg ] \,
,
\end{eqnarray}
where $\chi_{q_i}(q)$, $\chi_{\pm}(\sigma)$ are isospin and spin
functions, $\Lambda^{\pm}=\Omega_i \pm 1/2 \ge 0$. $r,z,\varphi$
are the cylindrical coordinates of ${\vec R}$.

The wave functions $\Phi_i$ are expanded into the eigenfunctions,
$\phi_{\alpha}$, of an axially symmetric deformed harmonic-oscillator
potential in cylindrical coordinates. We use 12 major
shells in this expansion,

\begin{equation}
\Phi_i \left( {\vec R},\sigma ,q\right) = \chi_{q_i}(q)
\sum_{\alpha} C^i_{\alpha} \phi_{\alpha}\left( {\vec R},\sigma
\right) \,  ,
\end{equation}
with $\alpha=\{n_r,n_z,\Lambda,\Sigma\}$ and
\begin{equation}
\phi_{\alpha}\left( {\vec R},\sigma \right)=
\psi^{\Lambda}_{n_r}(r) \psi_{n_z}(z) \frac{e^{i\Lambda
\varphi}}{\sqrt{2\pi}} \chi_{_\Sigma}(\sigma)\, ,
\end{equation}
in terms of Hermite and Laguerre polynomials
\begin{equation}
\psi_{n_z}(z)= \sqrt{\frac{1}{\sqrt{\pi}2^{n_z}n_z!}} \,
\beta^{1/2}_z\, e^{-{\xi}^2/2}\, H_{n_z}(\xi) \, ,
\end{equation}
\begin{equation}
\psi^{\Lambda}_{n_r}(r)=\sqrt{\frac{n_r}{(n_r+\Lambda )!}} \,
\beta_{\perp}\, \sqrt{2}\, \eta^{\Lambda/2}\, e^{-\eta/2}\,
L_{n_r}^{\Lambda}(\eta) \, ,
\end{equation}
with
\begin{eqnarray}
\beta_z=(m\omega_z/\hbar )^{1/2}&,&\quad
\beta_\perp=(m\omega_\perp/\hbar )^{1/2},\nonumber \\
\quad \xi=z\beta_z&,&\quad \eta=r^2\beta_\perp ^2 \, .
\end{eqnarray}

\subsection{Density distributions and root mean square radii}

The spin-independent proton and neutron densities are given by
\begin{equation}
\rho({\vec R})=\rho (r,z)=\sum _{i} 2v_i^2\rho_i(r,z)\, ,
\end{equation}
in terms of the occupation probabilities $v_i^2$ resulting from
the BCS equations and the single-particle densities $\rho_i$
\begin{equation}
\rho_i({\vec R})=  \rho_i(r,z)=|\Phi^+_i(r,z)|^2+
|\Phi^-_i(r,z)|^2 \, ,
\end{equation}
with
\begin{eqnarray}
\Phi^\pm _i(r,z)&=&{1\over \sqrt{2\pi}}\nonumber \\
&\times & \sum_{\alpha}\, \delta_{\Sigma, \pm 1/2}\,
\delta_{\Lambda,\Lambda^\mp}\, C_\alpha ^i\, \psi_{n_r}^\Lambda
(r) \, \psi_{n_z}(z) \, .
\end{eqnarray}
The multipole decomposition of the density can be written as
\cite{vautherin,oursmontse}
\begin{eqnarray}
\rho(r,z)& = &\sum_{\lambda} \rho_{\lambda}(R)
P_{\lambda}(\cos\theta)\nonumber \\
&=& \rho_0(R) + \rho_2(R)\,
P_2(\cos \theta) + \ldots \, , \label{rhomult}
\end{eqnarray}
with multipole components $\lambda$
\begin{eqnarray}
\rho_{\lambda}(R)&=&\frac{2\lambda +1}{2}
\int_{-1}^{+1} P_{\lambda}(\cos\theta)\nonumber \\
&\times & \rho(R\cos\theta,R\sin\theta) d(\cos\theta) \, ,
\end{eqnarray}
and normalization given by
\begin{equation}
\int \rho({\vec R}) d{\vec R} = X ;\qquad  4\pi \int
R^2dR\rho_0(R) = X \, ,
\end{equation}
with $X=Z,\, N$ for protons and neutrons, respectively.

The mean square radii for protons and neutrons are defined as
\begin{equation}
<r_{\rm p,n}^2> =\frac{ \int R^2\rho_{\rm p,n}({\vec R})d{\vec R}}
{\int \rho_{\rm p,n}({\vec R})d{\vec R}} \, , \label{r2pn}
\end{equation}
and the rms radii for protons and neutrons are simply given by
\begin{equation}
r_{\rm p,n}=<r_{\rm p,n}^2> ^{1/2} \, . \label{rmsrnp}
\end{equation}

The mean square radius of the charge distribution in a nucleus can
be expressed as
\begin{eqnarray}
<r^2_{\rm ch}>&=&<r^2_{\rm p}>+<r^2_{\rm ch}>_{\rm p} +(N/Z)
<r^2_{\rm ch}>_{\rm n}\nonumber \\
&+& r^2_{\rm CM} + r^2_{\rm SO}
\, , \label{rch}
\end{eqnarray}
where $<r^2_{\rm p}>$ is the mean square radius of the point
proton distribution in the nucleus (\ref{r2pn}), $<r^2_{\rm
ch}>_{\rm p}$ and $<r^2_{\rm ch}>_{\rm n}$ are the mean square
charge radii of the charge distributions in a proton and a
neutron, respectively. $r^2_{\rm CM}$ is a small correction due to
the center of mass motion, which is evaluated assuming
harmonic-oscillator wave functions. The last term $r^2_{\rm SO}$ is a tiny
spin-orbit contribution to the charge density. Correspondingly, we
define the charge rms radius
\begin{equation}
r_{\rm c}=<r_{\rm ch}^2> ^{1/2} \, . \label{rmsrc}
\end{equation}

\subsection{Form factors and diffraction parameters}

Besides the mean square radii, additional characteristics of the
density distributions can be deduced from the Fourier transforms
of these densities. The form factors are defined as
\begin{equation}
F_{\rm p,n}(\vec{q}) = \frac {\int \rho_{\rm p,n}({\vec R})e^{i
\vec{q} \cdot \vec{R}} d{\vec R}} { \int \rho_{\rm p,n} ({\vec
R})d{\vec R} }\, . \label{ff}
\end{equation}

In the Plane Wave Born Approximation (PWBA) the elastic electron
scattering cross sections are related to the Fourier transform of
the charge density
\begin{equation}
F_{\rm ch}(\vec{q}) = \frac{1}{Z} \int \rho_{\rm ch}({\vec R})e^{i
\vec{q} \cdot \vec{R}} d{\vec R} \, ,
\end{equation}
where $\vec{q}$ is the momentum transfer by the virtual photon in
the scattering process.

For each density multipole $\lambda$, one defines a $C\lambda$ form factor

\begin{equation}
F^{C\lambda} (q)= \frac{4\pi}{X}  \int_0^\infty
R^2dR \rho_{\lambda}(R) j_{\lambda} (qR)\, .
\end{equation}
In particular,
\begin{equation}
F^{C0} (q)= \frac{4\pi}{X} \int_0^\infty R^2dR \rho_0(R) j_0 (qR)
\end{equation}
has the limit at $q \rightarrow 0 $
\begin{equation}
F^{C0} (q)\; \rightarrow \;\frac{4\pi}{X}\int R^2 dR \rho_0(R) = 1
\, .
\end{equation}

Elastic and inelastic electron scattering have been extensively used
to extract the various multipoles of the charge density, which show up
in different transitions.
In particular, in even-even deformed nuclei, $F^{C0}$ (and correspondingly
$\rho_0$) show up in the elastic cross section, while  $F^{C2}$
(and hence $\rho_2$) show up in the inelastic cross section for the
transition  $0^+ \rightarrow 2^+$  between the
band-head and first excited rotational state \cite{Elvira,Berdi}.

In the next sections we will study the neutron skin thickness. We
will use first the difference between the neutron and proton rms
radii to characterize the different spatial extensions of neutron
and proton densities. But as already noticed \cite{Mizu2000}, the
rms radii (second moments of the densities) provide a very limited
description of the nucleon density distributions. A more effective
tool to analyze skins \cite{Mizu2000,Bertu2007} is the Helm model
\cite{Helm56,Sprung92}. This is a model that allows one to extract
from the form factor in a simple way the two main characteristics
of the density, a diffraction radius and a surface thickness. In
this model one describes the density by convoluting a hard sphere
(hs) density having diffraction radius $R_d$ with a gaussian of
variance $\sigma$,

\begin{equation}
\rho_{\rm Helm}(r;R_d,\sigma)=\rho_{\rm
hs}(r;R_d)*\rho_G(r;\sigma) \, ,
\end{equation}
where
\begin{equation}
\rho_{\rm hs}(r,R_d)= \frac{3X}{4\pi R_d^3}  \Theta(R_d-r),
\end{equation}
and
\begin{equation}
  \rho_G(r;\sigma) =(2\pi \sigma^2)^{-3/2} e^{(-r^2/2\sigma^2)}  .
\end{equation}
The corresponding Helm form factor is
\begin{eqnarray}
F_{\rm Helm}(q)&=&F_{\rm hs}(q;R_d)F_G(q;\sigma)\nonumber \\
&=&\frac{3}{qR_d}j_1(qR_d) e^{-\sigma^2 q^2/2} \, .
\end{eqnarray}

Now, the most prominent feature of the density distribution,
namely its extension, can be related to the first zero in the form
factor, this is the diffraction radius
\begin{equation}
R_d=4.49341/q_1  \, , \label{rd}
\end{equation}
where $q_1$ is the first zero of the form factor.

The nuclear surface width $\sigma$ can be related to the height of
the second maximum of the form factor located at $q_{\rm max}$:
\begin{equation}
 \sigma^2 = \frac{2}{q_{\rm max}^2} \ln \frac{3 j_1(q_{\rm max}R_d)}{R_dq_{\rm max}
F(q_{\rm max})} \, . \label{sigma}
\end{equation}
The variance $\sigma$ is related to the surface thickness $t$
(defined as the distance over which the density decreases from 90\%
to 10\% of the central value) by $t=2.54\; \sigma$. Moreover, the
surface thickness $t$ is also related to the diffuseness $a$ in
the two-parameter Fermi distribution, by $t=4a\ln 3 =4.39\; a$.

Taking into account that the second moment of a convoluted
distribution is given by the sum of the second moments of the two
single distributions, one gets the Helm rms radius
\begin{equation}
R_{\rm rms}^{\rm Helm}=\sqrt{\frac{3}{5} \left( R_d^2 + 5 \sigma
^2 \right)} \, .
\end{equation}

Taking out the factors $\sqrt{3/5}$, which relate the rms radii to
the radii of the equivalent uniform hard spheres, we define

\begin{equation}
R_{\rm hs}=\sqrt{5/3} <r^2>^{1/2} \,
\end{equation}
and
\begin{equation}
R_{\rm Helm}=\sqrt{5/3}\, R_{\rm rms}^{\rm Helm}=
\sqrt{R_d^2+5\sigma ^2} \, .
\end{equation}

From these definitions we construct the following neutron-proton
radius differences that will be used in the next sections
\begin{equation}
\Delta R_d=R_d (n)-R_d (p) \, , \label{drd}
\end{equation}

\begin{eqnarray}
\Delta R_{\rm hs}&=&R_{\rm hs}(n) -R_{\rm hs} (p) \nonumber \\
&=&\sqrt{5/3} \left[ <r_n^2>^{1/2} - <r_p^2>^{1/2}\right] \, ,
\label{drhs}
\end{eqnarray}

\begin{equation}
\Delta R_{\rm Helm} =R_{\rm Helm}(n) -R_{\rm Helm} (p) \, .
\label{drh}
\end{equation}

\section{Results and discussion}

\subsection{Root mean square radii and density distributions}

\begin{figure*}
\centering
\includegraphics[width=140mm]{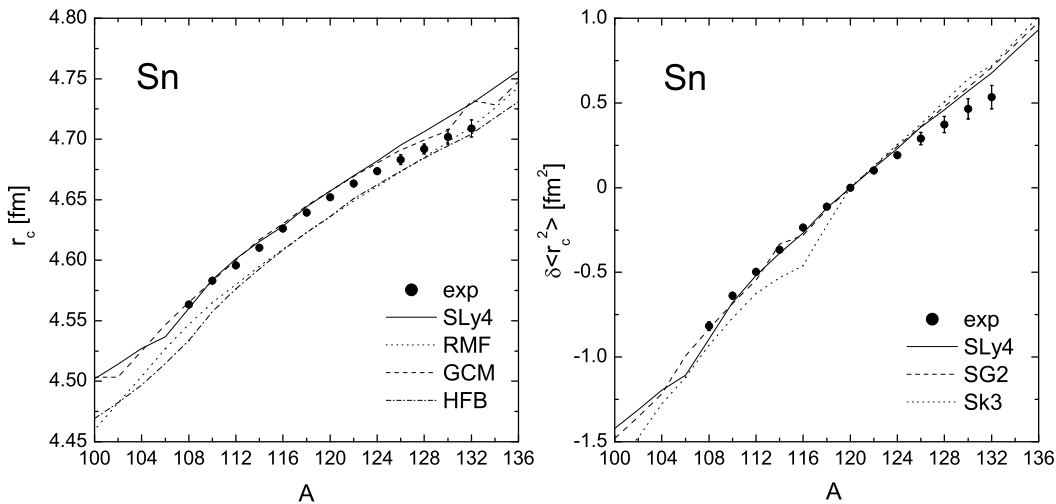}
\caption[]{Left panel: Charge rms radii $r_{c}$ of tin isotopes.
The SLy4 result is compared with the results from RMF calculations
\protect\cite{Lala99}, HFB \protect\cite{Libert2007} and GCM
\protect\cite{Libert2007}. Experimental data are from
\protect\cite{Blanc2002,Blanc2005,Ansel86,Piller90}; Right panel:
Theoretical (with different Skyrme forces) and experimental
isotope shifts $\delta\langle r_{c}^{2}\rangle $ of tin isotopes
relative to $^{120}$Sn. \label{fig1}}
\end{figure*}

We start by showing our results for the rms radii of the charge
distributions [Eq.~(\ref{rmsrc})]. We compare them to the
available experimental information obtained from various methods
including laser and muonic atoms spectroscopy
\cite{Blanc2002,Blanc2005,Ansel86,Piller90,Pomor95,fricke,Keim95}.
We also compare our results with different theoretical
calculations. They include RMF calculations with NL3
parametrization and pairing correlations in BCS approach (RMF in
Fig.~\ref{fig1}) \cite{Lala99}, nonrelativistic calculations
performed within HFB approach deduced under triaxial symmetry from
D1S Gogny effective interaction (HFB in Fig.~\ref{fig1}), as well
as calculations performed within a configuration mixing approach
in the space spanned by the constrained HFB states. The latter are
done within the GCM under Gaussian overlap approximation for the
complete quadrupole collective space (GCM in Fig.~\ref{fig1})
\cite{Libert2007}.

Beginning with Sn isotopes for which more data and calculations
are available, we show on the right panel of Fig.~\ref{fig1} our
results for the squared charge radii differences in Sn isotopes
obtained from three different Skyrme forces, SLy4, SG2 and Sk3. We
compare them to experiment, taking the radius of $^{120}$Sn as the
reference \cite{Piller90}. On the left panel we compare our SLy4
results for the charge radii with the other theoretical approaches
mentioned above. The general purpose of Fig.~\ref{fig1} is firstly
to show that different Skyrme forces do not differ much in their
predictions of charge rms radii and secondly, to show that our
results with SLy4 are comparable to other theoretical predictions
including approaches that go beyond the mean-field approximation,
as well as relativistic approaches. Then, by comparing our results
with experiment and with other theoretical results, we have
evaluated the quality of our calculations. We conclude that our
method reproduces the experimental data with a similar accuracy to
other microscopic calculations that, as explained above, may be
more sophisticated but may also be more time consuming. This 
agreement provides a good starting point to make predictions for
other quantities such as neutron-proton radii differences, where the
experimental information is scarce and it is not as accurate as in
the case of charge radii.

\begin{figure*}
\centering
\includegraphics[width=140mm]{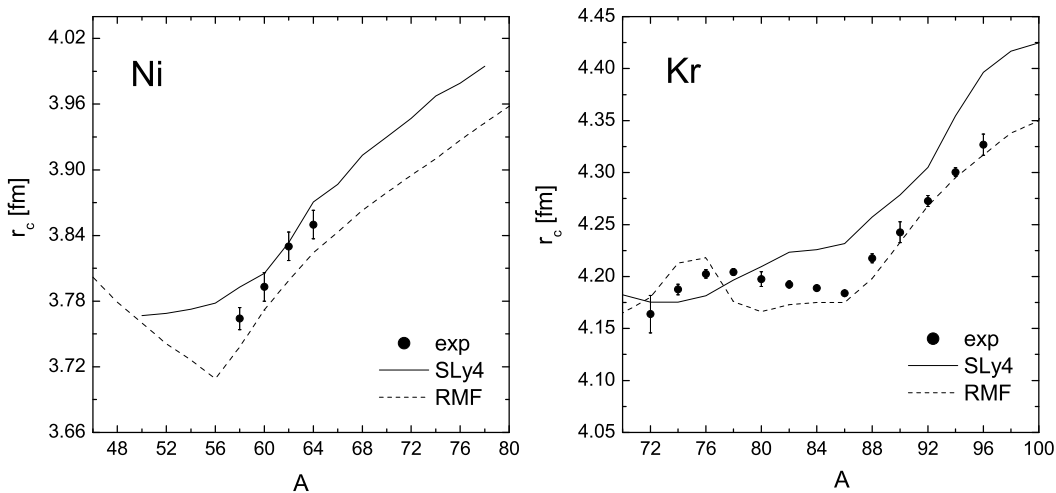}
\caption[]{Left panel: Charge rms radii $r_{c}$ of Ni isotopes.
The SLy4 results are compared with the results from RMF
calculations \protect\cite{Lala99}. Experimental data are from
\protect\cite{Pomor95,fricke}; Right panel: Charge rms radii
$r_{c}$ of Kr isotopes. The SLy4 results are compared with the
results from RMF calculations \protect\cite{Lala99}. Experimental
data are from \protect\cite{Keim95}. \label{fig2}}
\end{figure*}

We complete this comparison of charge radii in Fig.~\ref{fig2}. On
the left panel we show our results for Ni isotopes and compare
them with experiment \cite{Pomor95,fricke} and with results from
RMF calculations \cite{Lala99}. On the right panel we show the
same comparison for Kr isotopes. Data are taken from
\cite{Keim95}. In the Ni isotopes, we can see that the lower
values of the rms radii occur around the double magic nucleus
$N=Z=28$, and around the semi-magic $N=50$ in Kr isotopes. It is
also worth mentioning that the bump shown around $A=76$ in the RMF
calculations of Kr isotopes has its origin in the change of the
ground-state nuclear shape from oblate to prolate. In our case we
obtain a smooth line because we only consider oblate shapes in
this figure, as they correspond to the equilibrium shapes in most
cases.

\begin{figure*}
\centering
\includegraphics[width=80mm]{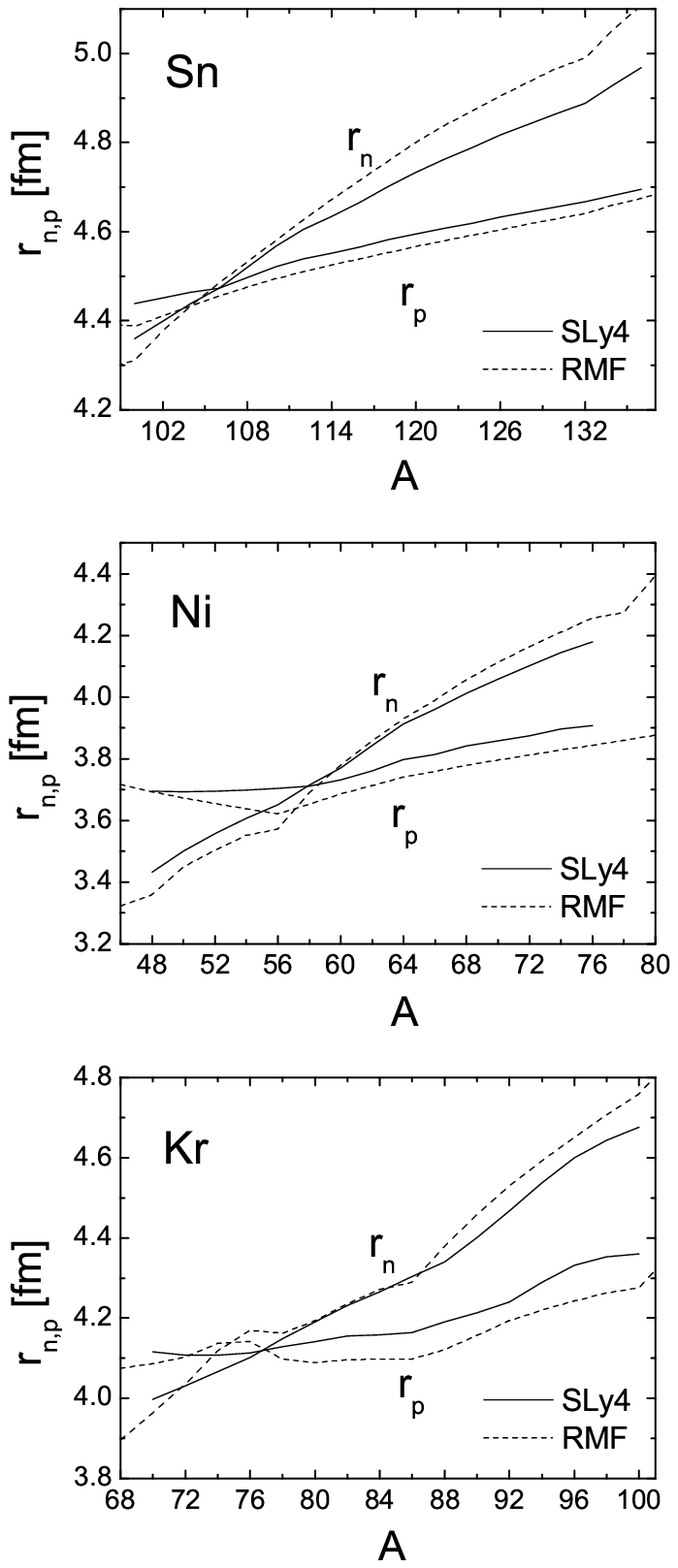}
\caption[]{Proton $r_{p}$ and neutron $r_{n}$ rms radii of Sn, Ni
and Kr isotopes calculated by using SLy4 force. The results from
RMF calculations \protect\cite{Lala99} are also given.
\label{fig3}}
\end{figure*}

\begin{figure*}
\centering
\includegraphics[width=140mm]{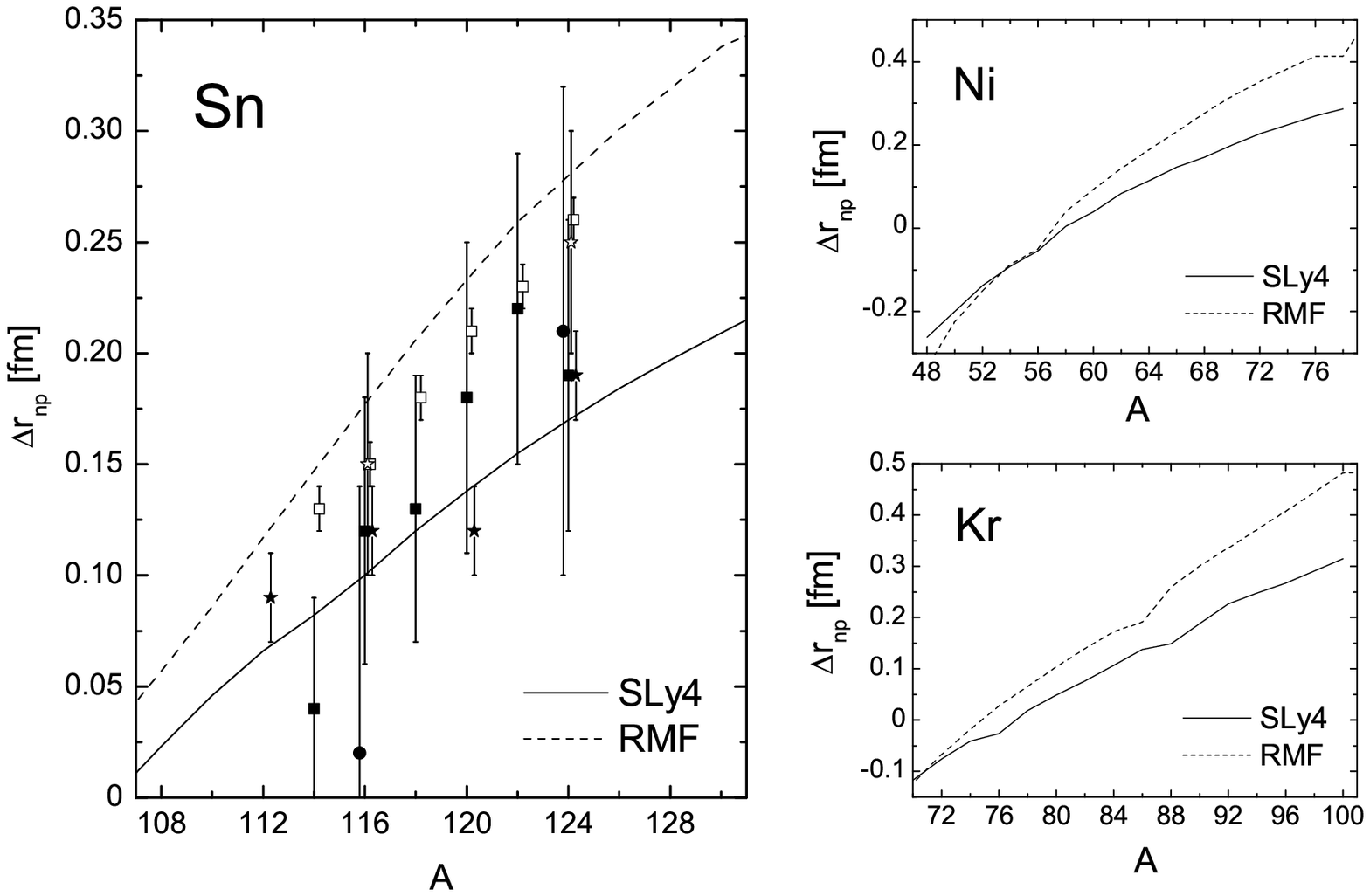}
\caption[]{Difference between neutron and proton rms radii $\Delta
r_{np}$ of Sn, Ni, and Kr isotopes calculated with SLy4 force. The
RMF calculation results are from Ref.~\protect\cite{Lala99}. The
experimental data for Sn isotopes measured in $(p,p)$ reaction
(open stars) \protect\cite{Ray79,Hoff81}, antiproton atoms (full
stars) \protect\cite{Agni2001}, giant dipole resonance method
(full circles) \protect\cite{Kras94} and spin dipole resonance
method (full and open squares) \protect\cite{Kras99,Kras2004} are
also shown. \label{fig4}}
\end{figure*}

Once we have confirmed that the agreement between our calculations
with the experimental $r_c$ radii is satisfactory, we have
guarantees that meaningful results will be obtained for the
neutron and proton mean square radii (\ref{rmsrnp}) by using the
same formalism with the same forces. Figure~\ref{fig3} contains
our results with the SLy4 force for those radii in the three
isotopic chains. They are compared with the predictions from RMF
\cite{Lala99}. We see that the tendency in the radii as a function
of the mass number $A$ is quite similar in both approaches, but in
general the proton rms radii with Skyrme are systematically larger
than the results from RMF. The situation is the opposite with
respect to the neutron rms radii. At the same time the latter
increase more slowly when calculated with SLy4. As a result we
will get systematically differences between the neutron and proton
rms radii, which are larger in the case of RMF as compared to the
case of Skyrme forces. This is clearly seen in Fig.~\ref{fig4}
where we plot the differences between the rms of neutrons and
protons $\Delta r_{np}=r_n-r_p$. On the left panel we show our
results for Sn isotopes and compare them to RMF results and to
experimental data taken from $(p,p)$ scattering
\cite{Ray79,Hoff81}, antiprotonic atoms \cite{Agni2001}, giant
dipole resonance method \cite{Kras94}, and spin dipole resonance
method \cite{Kras99,Kras2004}. As we can see in Fig.~\ref{fig4}
the experimental data are located between the predictions of both
theoretical approaches and in general, there is agreement with
experiment within the error bars. On the right panels we see the
predictions for $\Delta r_{np}$ in the cases of Ni and Kr
isotopes, where there are no data.

The RMF results for the difference $\Delta r_{np}$ systematically
overestimate the Skyrme HF results, as it can be seen from
Fig.~\ref{fig4}. The reason for this is related to the difference
in the nuclear symmetry energy and, consequently, to the different
neutron equation of state (EOS) which has been extensively studied
in recent years \cite{Yoshida2006,Chen2005,Diep2003,Diep2007}. It
was shown that there exists a linear correlation between the
derivative of the neutron EOS (or the pressure of neutron matter)
and the neutron skin thickness in heavy nuclei (defined as $\Delta
r_{np}=r_n-r_p$) in both Skyrme HF \cite{Brown2000,Furn2002} and
RMF \cite{Furn2002,Typel2001} models. We note that also a relation
between $\Delta r_{np}$ and both volume and surface
symmetry energy parameters was established recently by
Danielewicz \cite{Daniel2003} and Steiner {\it et al.}
\cite{Steiner2005} which provides a consistent description of
nuclei with neutron excess. Typel and Brown \cite{Typel2001}
demonstrated that the relativistic models produce larger neutron
radii compared with the nonrelativistic ones, reflecting the fact
that the saturation density of asymmetric matter is lower in the
EOS when phenomenological nucleon interaction in the RMF theory is
used \cite{Oyam98}. The results shown for neutron radii in Fig. 3
and correspondingly for neutron thicknesses in Fig. 4 support the
above general conclusion.

\begin{figure*}
\centering
\includegraphics[width=140mm]{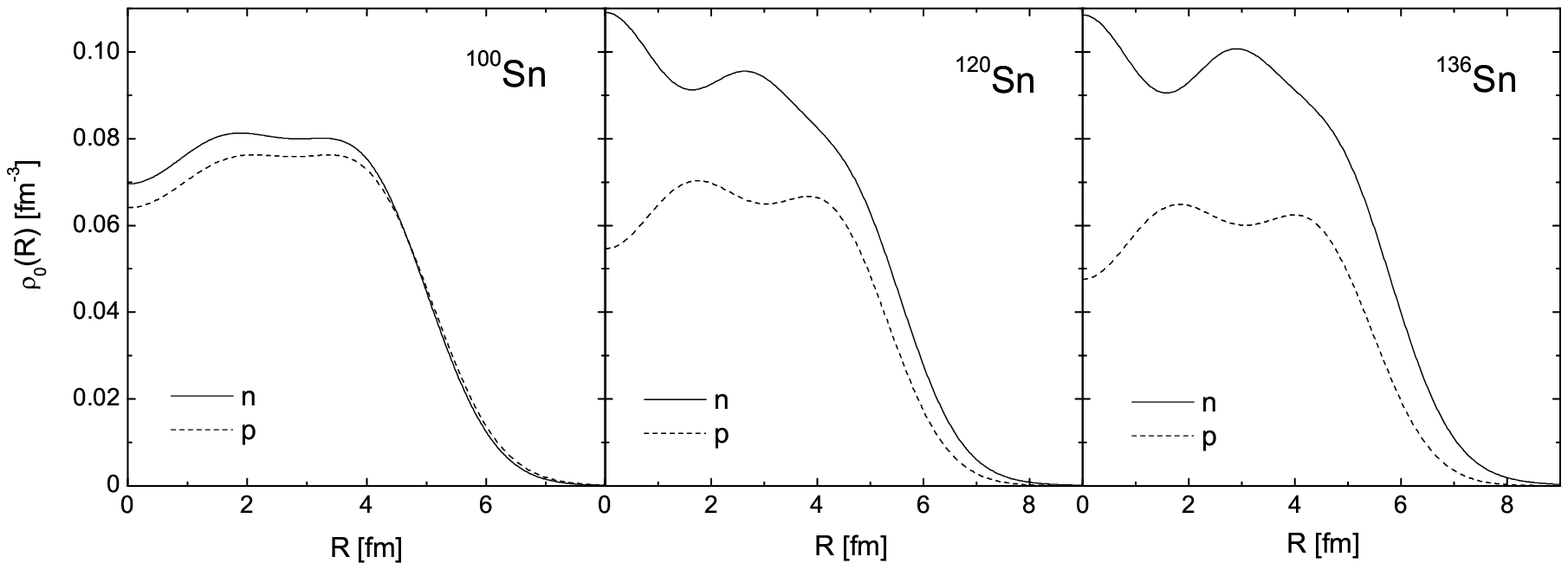}
\caption[]{HF+BCS proton and neutron densities $\rho_{0}(R)$ of
$^{100}$Sn, $^{120}$Sn, and $^{136}$Sn calculated with SLy4
force.
\label{fig5}}
\end{figure*}

\begin{figure*}
\centering
\includegraphics[width=140mm]{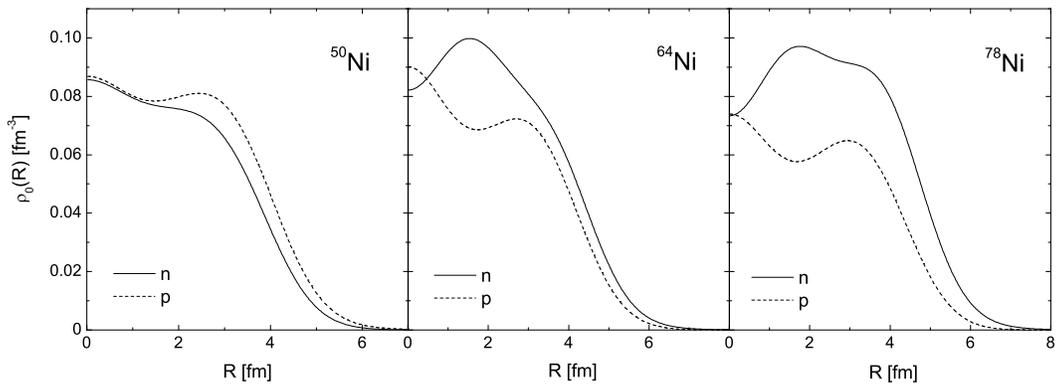}
\caption[]{Same as in Fig.~\ref{fig5}, but for $^{50}$Ni,
$^{64}$Ni, and $^{78}$Ni.\label{fig6}}
\end{figure*}

\begin{figure*}
\centering
\includegraphics[width=140mm]{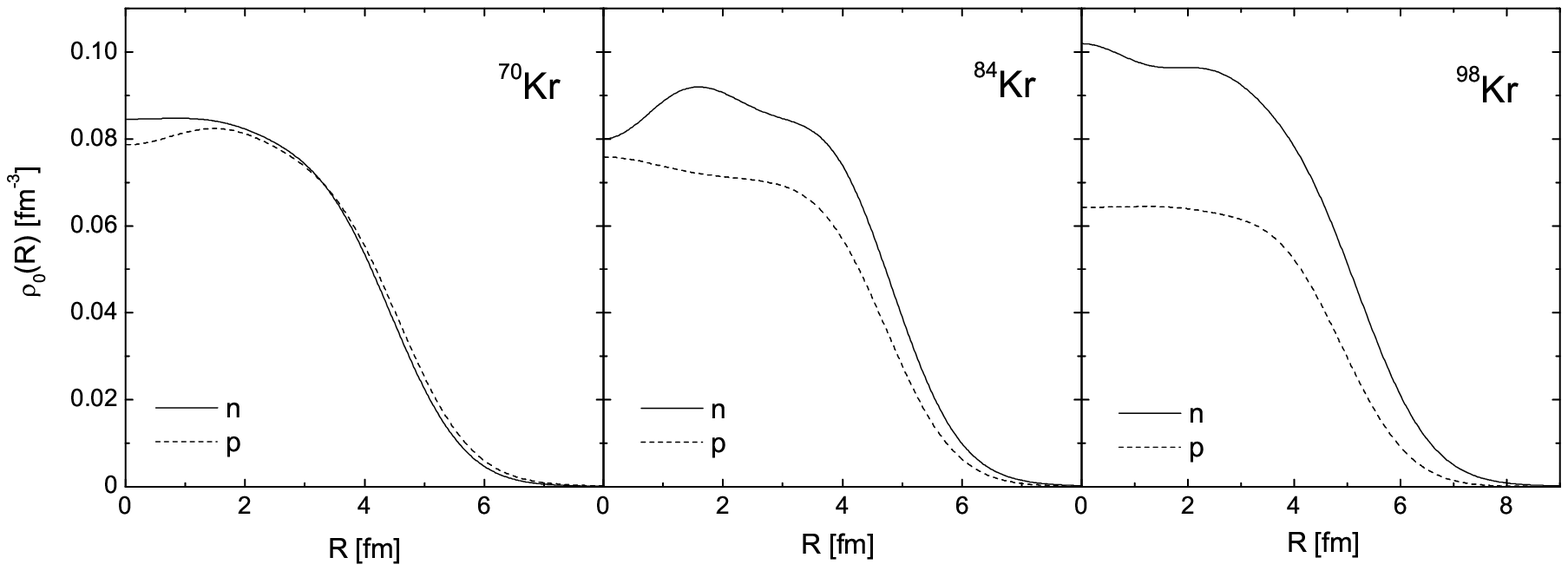}
\caption[]{Same as in Fig.~\ref{fig5}, but for $^{70}$Kr,
$^{84}$Kr, and $^{98}$Kr.\label{fig7}}
\end{figure*}

In the next figures we show the proton and neutron density
distributions  $\rho_0(R)$ (\ref{rhomult}) of some selected
isotopes in the three chains considered. We have chosen two
extreme neutron-deficient and neutron-rich isotopes and one stable
isotope between them. Figure~\ref{fig5} shows the neutron (solid)
and proton (dashed) densities in the $^{100,120,136}$Sn isotopes.
From left to right we see the evolution of these densities as we
increase the number of neutrons. In the case of $^{100}$Sn
($N$=$Z$=50) we see that the two densities are practically the
same except for Coulomb effects that make the protons to be more
extended and, therefore, this has to be compensated with a small
depression in the interior. The effect of adding more and more
neutrons is to populate and extend the neutron densities. This
makes also the proton distribution to follow the neutron one,
increasing its spatial extension. The cost of this radius enlargement in
the case of protons is a depression in the nuclear interior
to preserve the normalization to the constant number of protons
$Z=50$. Then, it can be seen graphically the emergence of a region
at the surface where the protons have practically disappeared
while the neutrons still survive. We will quantify later this
region in terms of the neutron skin thickness definitions.
Figures~\ref{fig6} and \ref{fig7} show the same information as in
Fig.~\ref{fig5} but for $^{50,64,78}$Ni and $^{70,84,98}$Kr
isotopes, respectively. The behavior of these densities corroborates
the comments made on the case of Sn isotopes.

\begin{figure*}
\centering
\includegraphics[width=140mm]{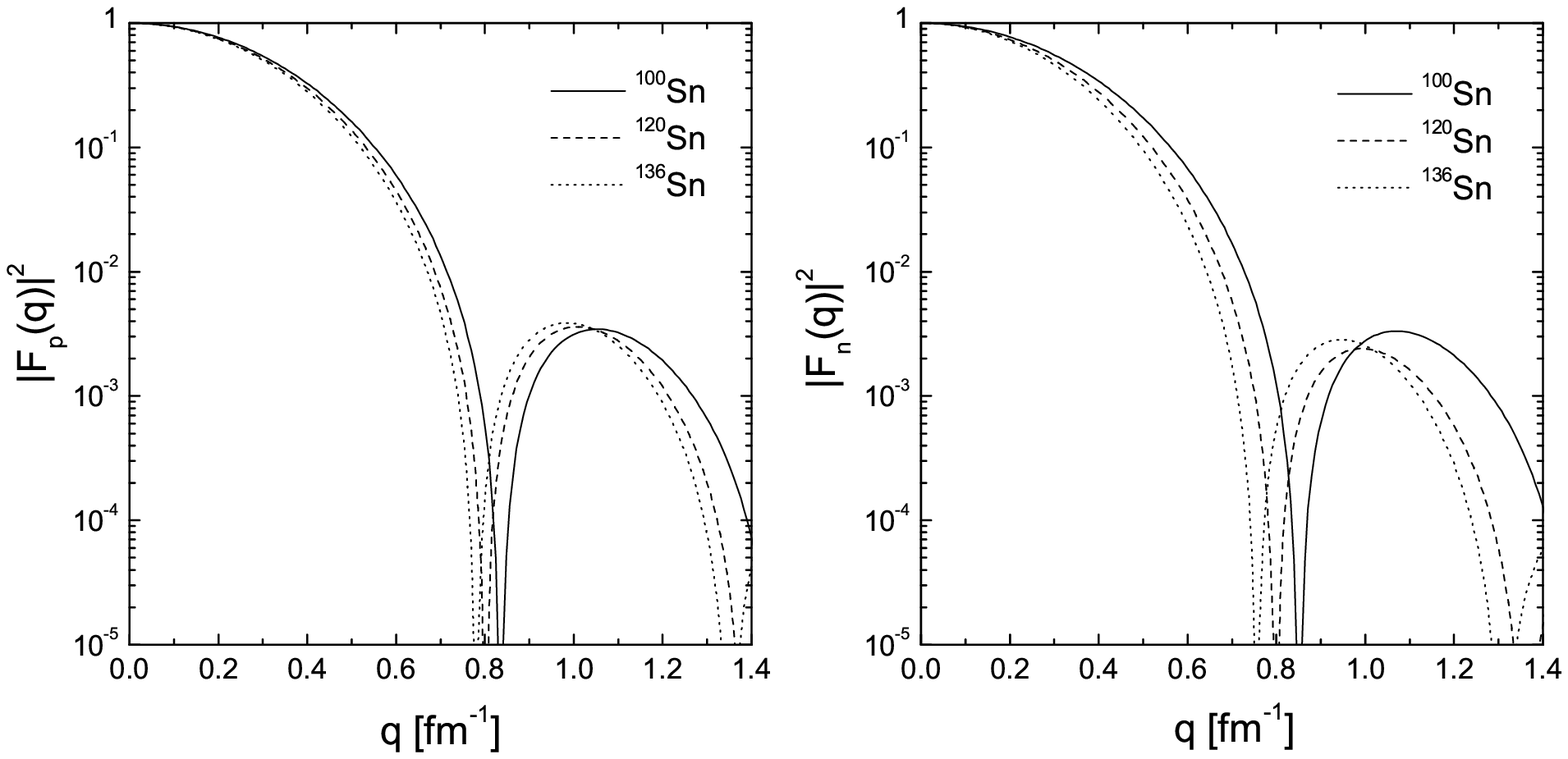}
\caption[]{Proton (left panel) and neutron (right panel) form
factors for the $^{100}$Sn, $^{120}$Sn, and $^{136}$Sn isotopes
are calculated in the PWBA.\label{fig8}}
\end{figure*}

As we mentioned in the last chapter, we will also characterize the
skin thickness in terms of diffraction parameters $R_d$ and
$\sigma$ deduced from the form factors. Figure~\ref{fig8} contains
these form factors (\ref{ff}) for protons
and neutrons of the three Sn isotopes shown in Fig.~\ref{fig5}. We
can see how the diffraction zeroes at $q_1$ (\ref{rd}) and the
location and magnitude of the second maximum, $q_{\rm max}$ and
$F(q_{\rm max})$ (entering in Eq.~(\ref{sigma})) needed to extract
$R_d$ and the surface width $\sigma$, change with the neutron
number. Thus, we see that the $q_1$ values diminish with
increasing neutron number and therefore $R_d$ increases
accordingly for both protons and neutrons. The values of
$q_{\rm max}$ are also reduced when $A$ increases but the values
of the form factor at these  $q_{\rm max}$ are rather similar.
Consistently, the parameters $\sigma$
extracted from (\ref{sigma}) are fairly similar.

\subsection{Neutron skin thickness}

The thickness of a neutron skin in nuclei may be defined in
different ways. One of these possibilities is to define it as the
difference between the root mean square radius of neutrons and
that of protons, as we have plotted in Fig.~\ref{fig4}. Similarly,
it can be defined as the difference between the neutron and proton
radii of the equivalent uniform spheres [Eq.~(\ref{drhs})].
Alternatively, it can be defined as the  difference between the
neutron and proton diffraction radii (\ref{drd}) or Helm radii
(\ref{drh}). All of these quantities have already been discussed
and used in the past as possible ways to quantify the skin
thickness (see for example \cite{Mizu2000}), arriving to the
conclusion that the radii difference defined in (\ref{drhs})
contains contribution from halo effects and the radii difference
defined in  (\ref{drh}) is a better measure of the skin.
Nevertheless, qualitatively the difference between the two definitions becomes
only apparent when dealing with very neutron-rich isotopes, which
are presently beyond the experimentally observed isotopes and out
of the scope of this paper.

On the other hand, the skin thickness can be also defined in terms
of some criteria that the neutron and proton densities must
fulfill. In Ref.~\cite{Fuku93} the neutron skin thickness is
defined as the difference between two radii, $R_1$ and $R_2$.
$R_1$ is the radius at which the ratio of the neutron density to the proton
density is equal to some given value (4 in \cite{Fuku93}). $R_2$ is
the radius at which the neutron density becomes
smaller than some percentage of the density at the center of the
nucleus (1 \% in \cite{Fuku93}). When this difference, $\Delta
R=R_2 - R_1$, is larger than some established value (in
\cite{Fuku93} this value is 1 fm, which is comparable to the range
of the nuclear force), a neutron skin with skin thickness  $\Delta
R$ is said to occur.

The factors used to define the skin thickness in the above
criteria could have been differently chosen in rather arbitrary
manners. Therefore, the absolute sizes of the skin thickness do
not have a very precise meaning. Nevertheless, these values are
useful to judge how the nucleon skins develop as the number of
nucleons change. Indeed, we have also considered the case where
the first criterion for the inner radius $R_1$ of the neutron skin
is changed. We use instead of the above criterion for $R_1$, the
radius at which the proton density becomes smaller than 1\% of the
latter at the center, which is similar to the criterion used to
define the outer radius $R_2$, but in this case for proton density
instead of the neutron density. When we use the conditions in Ref.
\cite{Fuku93}, we call it criterion (a). When we use the
alternative condition for $R_1$, we call it criterion (b).

\begin{figure*}
\centering
\includegraphics[width=140mm]{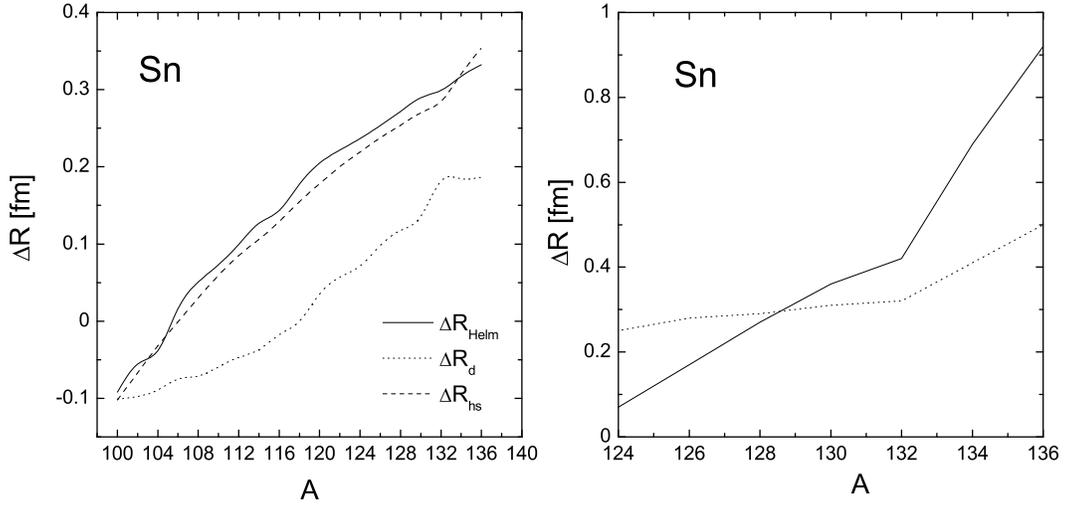}
\caption[]{Neutron skin thicknesses for tin isotopes. Left panel:
$\Delta R_{\rm d}$ [Eq.~(\ref{drd})], $\Delta R_{\rm hs}$
[Eq.~(\ref{drhs})], and $\Delta R_{\rm Helm}$ [Eq.~(\ref{drh})];
Right panel: corresponding to criterion (a) (solid line) and
criterion (b) (dotted line). \label{fig9}}
\end{figure*}

We show in Fig.~\ref{fig9} the results obtained for the neutron
skin thickness in Sn isotopes according to the different
definitions discussed above. The left panel contains the results
for definitions involving directly the difference between neutron
and proton radii, either the equivalent hard spheres radii $\Delta
R_{\rm hs}$ [Eq.~(\ref{drhs})] corresponding to the rms radii, the
diffraction radii $\Delta R_{\rm d}$ [Eq.~(\ref{drd})], and the
Helm radii $\Delta R_{\rm Helm}$ [Eq.~(\ref{drh})]. The skin thickness
predicted by the
difference of the very simple diffraction radii is in general
smaller than the thickness predicted by the other two more
involved options that are very similar in this range of masses.
The right panel contains the neutron skin thickness defined
according to the criteria on the density distributions (a) (solid
line) and (b) (dashed line). They only differ in the way in which the
starting radius of the skin $R_1$ is chosen. One can see that we
obtain larger neutron skin thicknesses when using criterion (b) in
the lighter isotopes, but this is reversed for heavier isotopes
and we get larger thickness when using criterion (a). This fact is
confirmed also by the values of the radii $R_{1}$ and $R_{2}$ and
their differences $\Delta R$ listed in Table~\ref{table1} for the heaviest
three isotopes in each chain considered. In general, the formation
of a skin when using (a) starts at distances smaller than those in
case (b) or comparable with them, which leads to larger absolute
size of the neutron skin produced by criterion (a). It is in this
region of heavier isotopes where we can properly talk about a
neutron skin formation. In this region, criterion (b) somehow establishes
a lower limit for the skin thickness. The latter can be
arbitrarily enlarged by relaxing the $\rho_n/\rho_p$ condition to values lower
than 4. Similar comments apply also to the next figures,
Fig.~\ref{fig10} for Ni isotopes and Fig.~\ref{fig11} for Kr
isotopes.

\begin{figure*}
\centering
\includegraphics[width=140mm]{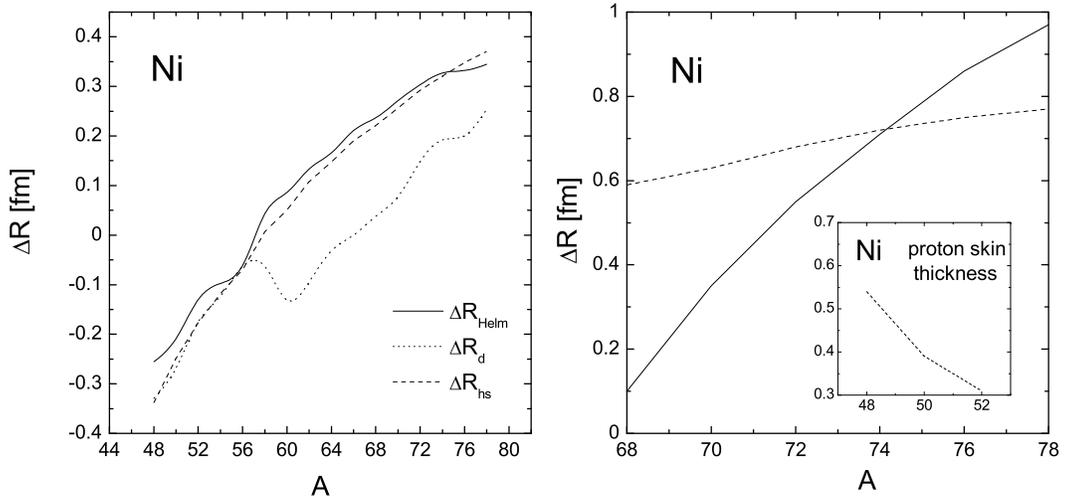}
\caption[]{Same as in Fig.~\ref{fig9}, but for Ni isotopes. A
formation of proton skin thickness with the criterion (b) is also
shown.\label{fig10}}
\end{figure*}

\begin{figure*}
\centering
\includegraphics[width=140mm]{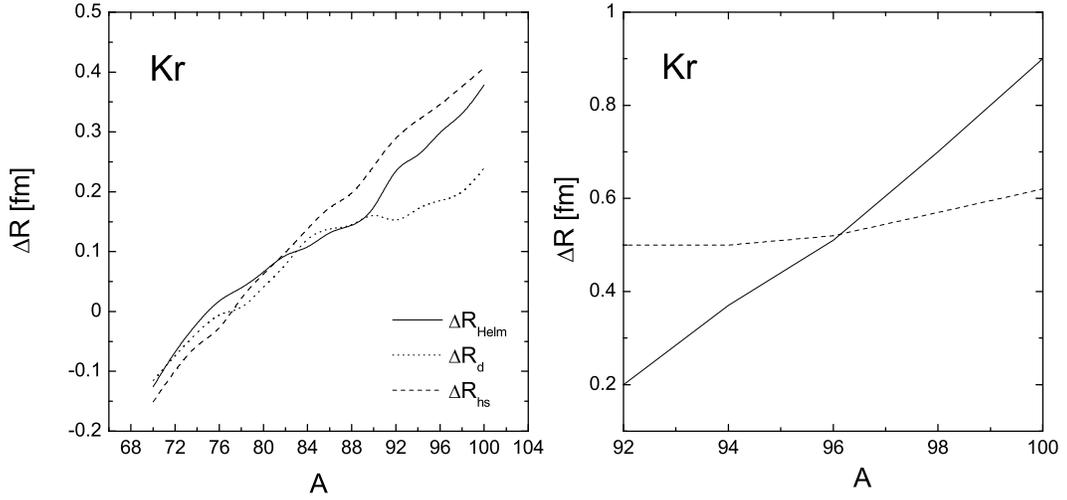}
\caption[]{Same as in Fig.~\ref{fig9}, but for Kr
isotopes.
\label{fig11}}
\end{figure*}

\begin{table}[h]
\caption{Radii $R_{1}^{(a)}$, $R_{1}^{(b)}$, $R_{2}$ and their
differences (skin thicknesses) $\Delta R^{(a)}=R_{2}-R_{1}^{(a)}$
and $\Delta R^{(b)}=R_{2}-R_{1}^{(b)}$ (in fm) according to
criteria (a) and (b).} {\begin{tabular}{ccccccccccccc} \hline
\hline Nuclei & & & $R_{1}^{(a)}$ & & $R_{1}^{(b)}$ & & $R_{2}$ &
&
$\Delta R^{(a)}$ & & $\Delta R^{(b)}$\\
\hline
$^{74}$Ni     & & & 6.49 & & 6.48 & & 7.20 & & 0.71 & & 0.72\\
$^{76}$Ni     & & & 6.38 & & 6.49 & & 7.24 & & 0.86 & & 0.75\\
$^{78}$Ni     & & & 6.32 & & 6.52 & & 7.29 & & 0.97 & & 0.77\\
$^{96}$Kr     & & & 7.32 & & 7.31 & & 7.83 & & 0.51 & & 0.52\\
$^{98}$Kr     & & & 7.20 & & 7.33 & & 7.90 & & 0.70 & & 0.57\\
$^{100}$Kr    & & & 7.06 & & 7.34 & & 7.96 & & 0.90 & & 0.62\\
$^{132}$Sn    & & & 7.64 & & 7.76 & & 8.08 & & 0.44 & & 0.32\\
$^{134}$Sn    & & & 7.50 & & 7.78 & & 8.19 & & 0.69 & & 0.41\\
$^{136}$Sn    & & & 7.40 & & 7.82 & & 8.32 & & 0.92 & & 0.50\\
\hline \hline
\end{tabular}}
\label{table1}
\end{table}

We would like to emphasize that although different definitions of
the neutron skin thickness produce different absolute values for
it, the relative skin thicknesses corresponding to the evolution
as the number of neutrons increase indicates the formation of such
a skin that can be expected to start at $A>132$ in Sn, $A>74$ in
Ni, and $A>96$ in Kr isotopes, as it is observed in
Figs.~~\ref{fig9}--\ref{fig11}.

Finally, we also consider the most neutron-deficient region of Ni
isotopes in a search for the formation of a proton skin. We have
already seen in the left panel in Fig.~\ref{fig10} that the
neutron skin thickness defined in terms of differences between
neutron and proton radii become negative at some point, indicating
that the proton distribution extends beyond the neutron one. This
can be further explored by reversing the definitions of $R_1$ and
$R_2$ interchanging the role of protons and neutrons. We show the
results in the inset of the right panel in Fig.~\ref{fig10}, where
we have applied the criterion (b) with protons and neutrons
interchanged. We find no proton skin when applying criterion (a).
One can see that a small skin starts developing in
these isotopes but we cannot push it further because $^{48}$Ni is
already at the proton drip line. The results are then not conclusive
enough to assess the existence of a proton skin in these isotopes.
This possibility
could be explored in the future in the most proton-rich nuclei
approaching the proton drip lines of lighter nuclei with $Z>N$.

\subsection{Neutron skin and deformation}

\begin{figure*}
\centering
\includegraphics[width=140mm]{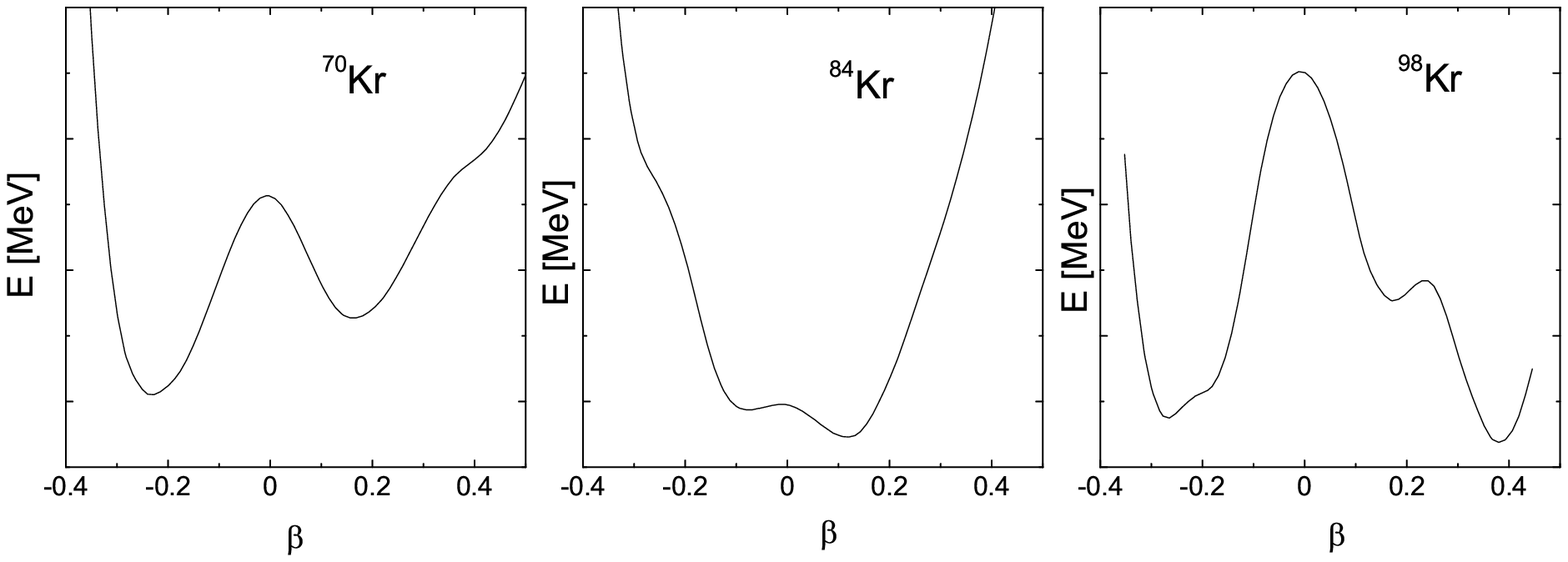}
\caption[]{Binding energies $E$ calculated with the SLy4 force as
a function of the quadrupole parameter $\beta$ for the even-even
$^{70}$Kr, $^{84}$Kr, and $^{98}$Kr isotopes. \label{fig12}}
\end{figure*}

When the nucleus is deformed, the thickness of the neutron skin
might depend on the direction. It is an interesting and natural
question to ask whether the deformed densities give rise to a
different skin size in the different directions.
It is also interesting to know whether the emergence of the skin may be
influenced by the nuclear shape.
We study in this work
such a dependence on the example of Kr isotopes, which are
examples of well deformed nuclei characterized by a large variety
of competing nuclear shapes \cite{constraintourskr}. Constraint HF+BCS
calculations \cite{constraint,constraintourskr} show also the
possibility of shape coexistence in these nuclei. The results
which we obtain for the binding energy of the three previously
selected Kr isotopes as a function of the quadrupole parameter
$\beta =\sqrt{\pi/5}Q_p/(Zr^2_p)$ ($Q_{p}$ being the proton
quadrupole moment) are presented in Fig.~\ref{fig12}. In this
figure the distance between two ticks in the vertical axis is
always 1 MeV but the origin is different for each curve. As we can
see, both prolate and oblate shapes produce minima very close in
energy. Then, we have chosen the neutron rich isotope $^{98}$Kr to
study the sensitivity of the neutron skin thickness to the various
directions in the two shapes.

GCM calculations built on the constrained HF+BCS
states may be carried out in order to describe more properly some
ground-state properties in deformed nuclei. In the
case of $^{98}$Kr the potential energy curve (Fig.~\ref{fig12})
shows pronounced minima at oblate and prolate shapes, which
are separated by an energy barrier of about 6 MeV. Thus, one
expects the ground state of $^{98}$Kr to be basically described by
a linear combination of these two configurations.

\begin{figure*}
\centering
\includegraphics[width=80mm]{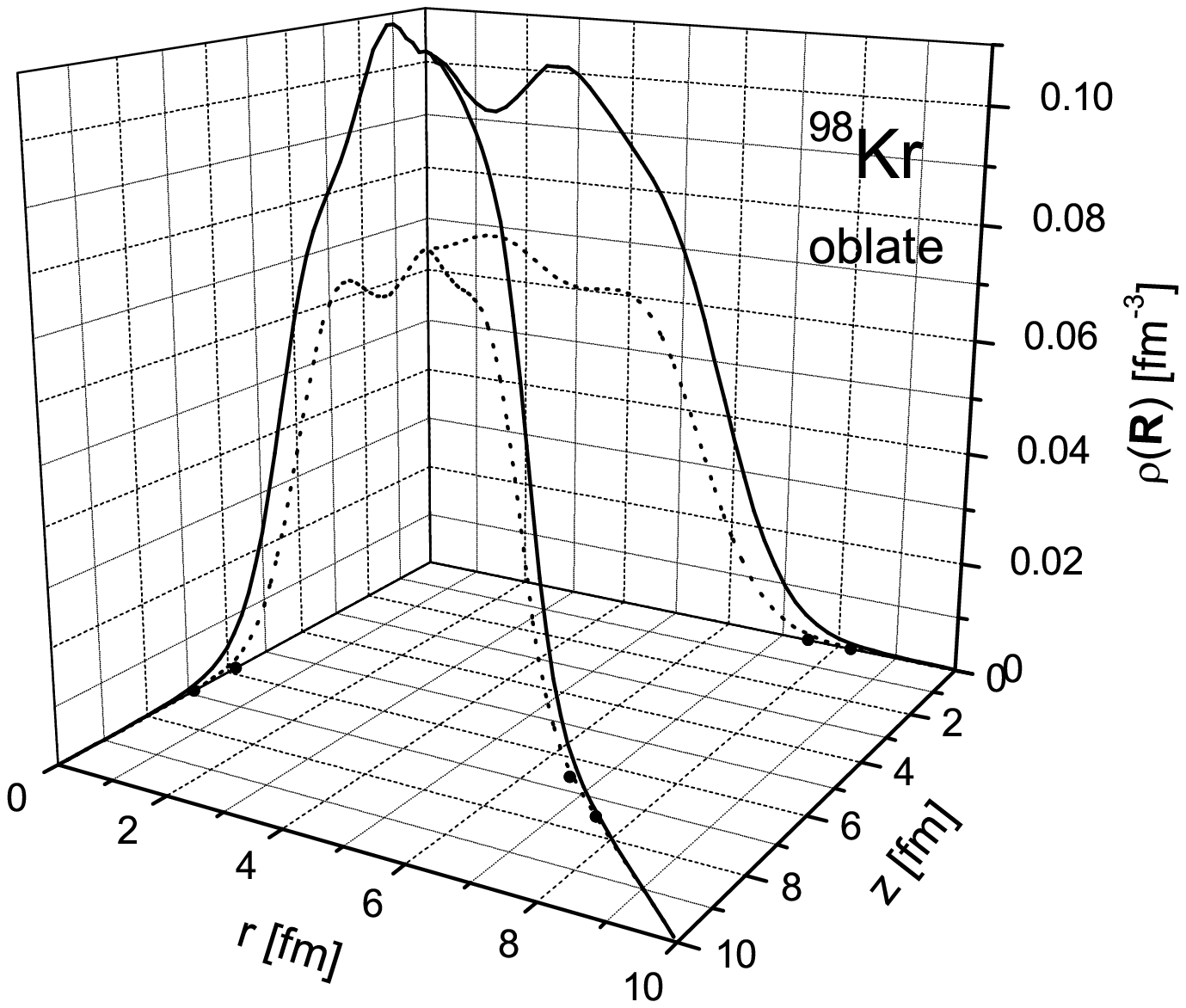}
\caption[]{Neutron (solid line) and proton (dotted line) density
distributions $\rho(\vec R)$ in different directions for oblate
shape of $^{98}$Kr. The full dots shown on the $(r,z)$ plane
correspond to radii $R_{1}$ and $R_{2}$ according to criterion
(a).\label{fig13}}
\end{figure*}

\begin{figure*}
\centering
\includegraphics[width=80mm]{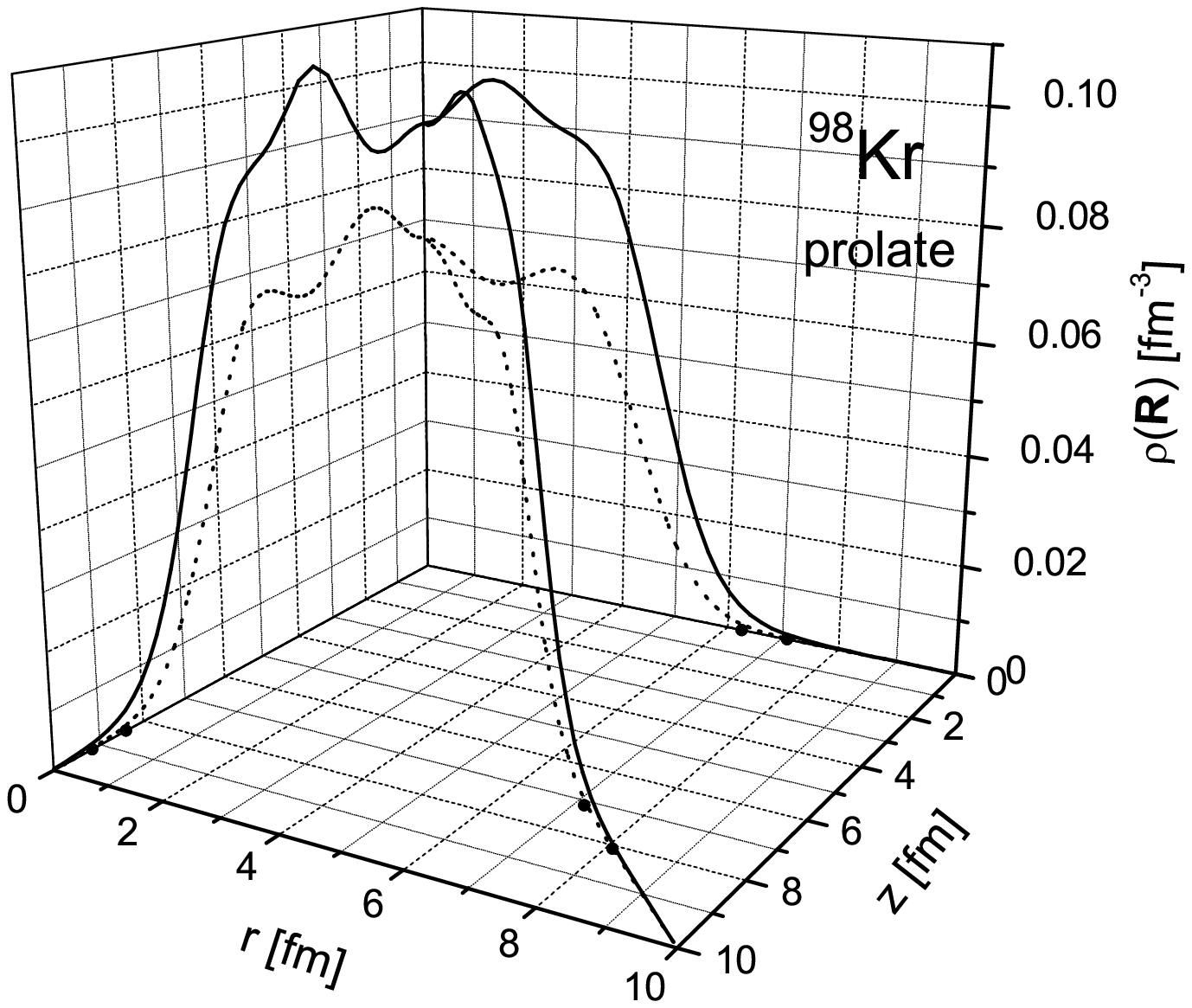}
\caption[]{Same as in Fig.~\ref{fig13}, but for prolate shape of
$^{98}$Kr.
\label{fig14}}
\end{figure*}

We first study the intrinsic density distributions $\rho(\vec{R})$
in various selected directions. For that purpose we show in
Figs.~\ref{fig13} and \ref{fig14} the densities of $^{98}$Kr for
oblate and prolate shapes, respectively. We can see the spatial
distributions for neutrons (solid) and protons (dotted) in three
different directions: $z$-direction $(r=0)$, $r$-direction
$(z=0)$, and $r=z$ direction. We can observe that the profiles of
the densities as well as the spatial extensions change with the
direction. Clearly, the densities are more extended in the
$z$-direction in the case of prolate shapes. The opposite is true
in the case of oblate shapes. The case $r=z$ gives always
intermediate densities. We have added in the three directions a
couple of full dots, indicating the radii $R_1$ and $R_2$ that
defines the skin thickness according to the above mentioned
criterion (a).

\begin{figure*}
\centering
\includegraphics[width=140mm]{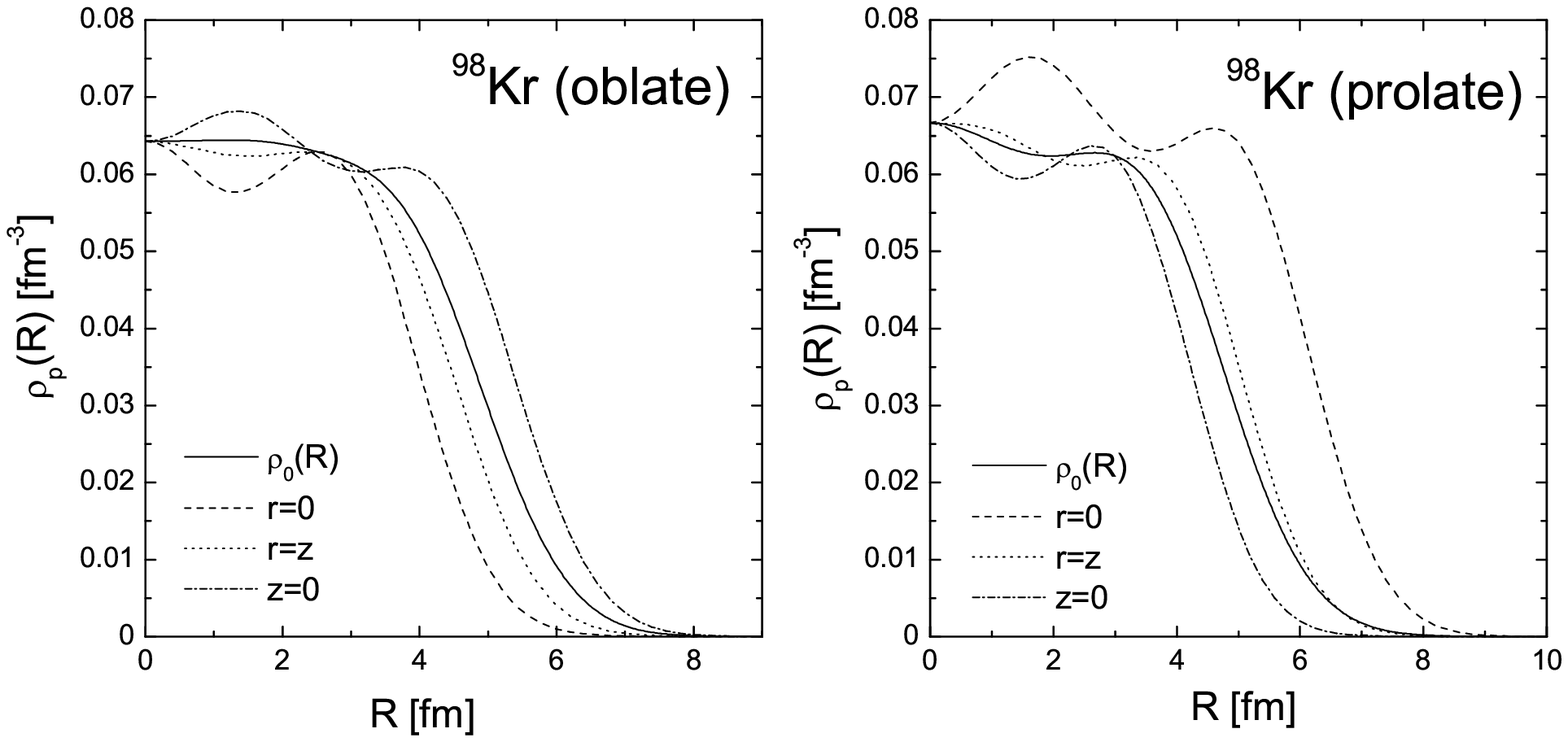}
\caption[]{Proton density distributions $\rho_{p}(R)$
corresponding to different directions for oblate (left panel) and
prolate (right panel) shape of $^{98}$Kr. The monopole component
$\rho_{0}(R)$ is also shown.
\label{fig15}}
\end{figure*}

The dependence of the intrinsic density on the different
directions can be also seen in Fig.~\ref{fig15}, where we plot as
an example the proton densities in the three directions mentioned
above for oblate (left) and prolate (right) shapes in the same
plane. We see more clearly  how the extension of the density in
the $z$-direction (labeled $r=0$) is the largest for the prolate
shape and the shortest for the oblate shape. We also plot for
comparison the monopole component $\rho_0 (R)$ (\ref{rhomult})
that lies between the two extreme cases and it is close to the
density in the $r=z$ direction.

\begin{figure*}
\centering
\includegraphics[width=80mm]{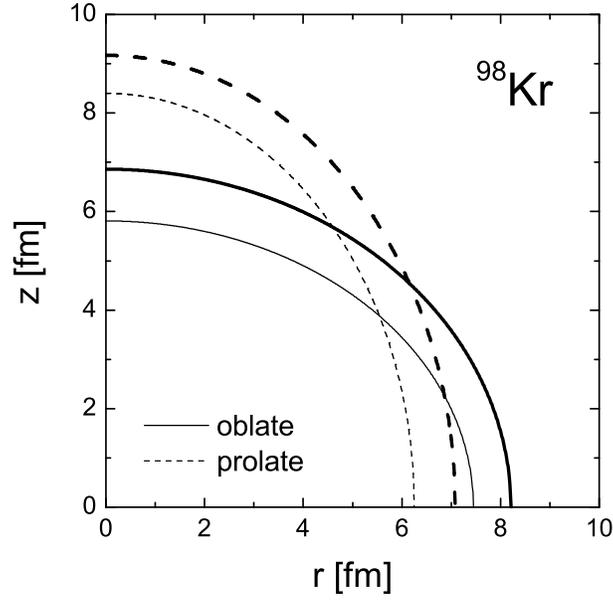}
\caption[]{Radii $R_{1}$ and $R_{2}$ according to criterion (a)
for neutrons (thick lines) and protons (thin lines) in $^{98}$Kr
nucleus (shown in $rz$ plane) corresponding to its oblate (solid
lines) and prolate (dashed lines) shape. \label{fig16}}
\end{figure*}

It is also worth looking at the points in the $(r,z)$ plane that
define the ellipses where the criteria for $R_1$ and $R_2$ are
met. Figure~\ref{fig16} shows these points for protons (thin
lines) and neutrons (thick lines) and for the two shapes, prolate
(solid) and oblate (dashed).
We can see that the size of the skin changes little with the directions
perpendicular to the surface,
but shows a tendency to increase on the shorter axis. It is interesting
to note that the skin size of the spherical component $\rho_0(R)$ is
an intermediate value. The overall skin thickness is also similar in
the oblate and prolate equilibrium shapes.
From this example we may conclude that the skin thickness does not depend
much on the oblate or prolate character of the deformation.
This is in line with the conclusions reached in Ref.~\cite{hamamoto} on
the example of Dy isotopes, where it was shown that the neutron
skin is nearly independent of the size of deformation
(spherical, deformed or superdeformed).

\begin{figure*}
\centering
\includegraphics[width=140mm]{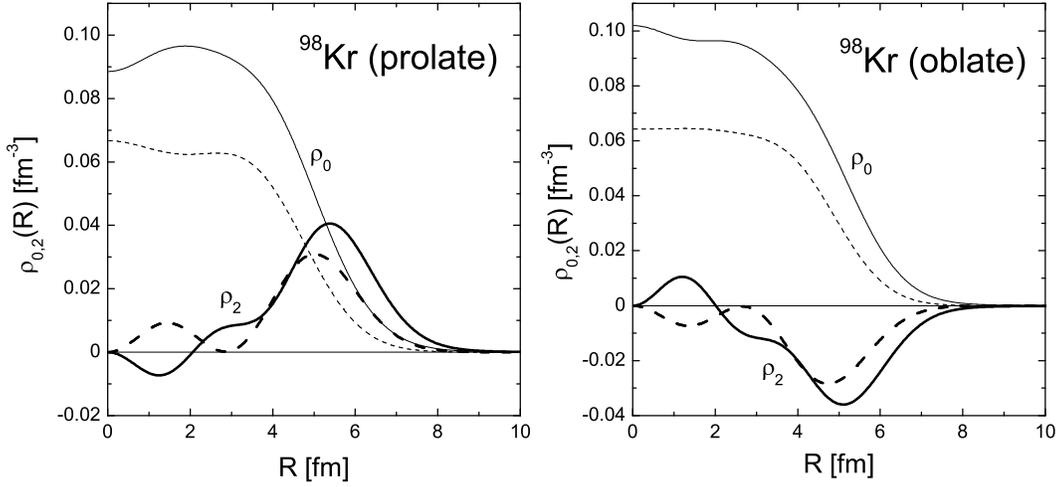}
\caption[]{Monopole $\rho_{0}(R)$ (thin lines) and quadrupole
$\rho_{2}(R)$ (thick lines) neutron (solid lines) and proton
(dashed lines) density distributions of $^{98}$Kr with oblate and
prolate shape.
\label{fig17}}
\end{figure*}

Figure~\ref{fig17} shows the monopole, $\rho_0(R)$, and quadrupole,
$\rho_2(R)$, components of the intrinsic density $\rho(\vec{R})$
(\ref{rhomult}) for protons (dashed lines) and neutrons (solid
lines) and for the oblate and prolate shapes in $^{98}$Kr. We can
see that $\rho_2(R)$ is peaked at the surface positively in the
case of the prolate deformation and negatively in the case of the
oblate one. This makes the total density in the $z$-direction to
be incremented with respect to the $\rho_0$ density in the prolate
case and to be decreased in the oblate one. The opposite is true
with respect to the direction perpendicular to the symmetry axis
$z$. We can also see that the skin thickness derived from the
$\rho_0$ components is quite similar to the thickness derived from
the quadrupole components $\rho_2$. This explains the
approximately constant skin thickness observed in the different
directions in Fig.~\ref{fig16}.

\section[]{CONCLUSIONS\label{s:concl}}

In this work we perform a theoretical analysis of nuclear skins,
exploring various definitions. For this purpose we
examine three chains of Ni, Kr, and Sn isotopes which might be of
particular interest in the future experiments in GSI and RIKEN.
The densities of these nuclei are calculated within a deformed
HF+BCS approach with Skyrme-type density-dependent effective
interactions \cite{constraintourskr}. We have shown
that this model gives a very reasonable description of the charge
rms radii of the Sn, Ni, and Kr isotopes and of the differences
between neutron and proton rms radii for several Sn isotopes. This
is confirmed by the good agreement with the available experimental
data, as well as with other theoretical predictions. Three Skyrme
parametrizations have been involved in the calculations: SG2, Sk3
and SLy4. Most of the results shown in the paper are obtained with
SLy4 force, but the other Skyrme interactions produce similar
results.

For the first time the various definitions
which have been previously proposed to determine the neutron skin
thickness, involving both matter radii and tails of nuclear
densities, have been compared within a deformed Skyrme HF+BCS model.
We find that all definitions of the neutron skin predict to a
different extent the existence of a skin in nuclei far from the
stability line. Particularly, a pronounced neutron skin can be
attributed to heavier isotopes of the three chains considered,
namely with $A>132$ for Sn, $A>74$ for Ni, and $A>96$ for Kr
isotopes. We also find that for a given isotopic chain
the increase of the skin with the neutron number in the
neutron-rich nuclei exhibits a rather constant slope, which is
different depending on the definition of nuclear skin. More
significant neutron skin is obtained when analyzing its formation
by means of definition from Ref.~\cite{Fuku93} (called criterion
(a)) or using an alternative one (called criterion (b)). In this
case we get an absolute size of the skin larger than 0.4 fm and
almost reaching 1 fm for the heaviest isotopes (in the case of
criterion (a)). At the same time, the neutron skin determined by
the difference between neutron and proton radii using diffraction
parameters defined in the Helm model shows a more smooth gradual
increase with the neutron excess and it is in size of around
0.3--0.4 fm. We would like to note that our results for Sn
isotopes are consistent with the results of calculations from
Ref.~\cite{Mizu2000} with SLy4 parametrization. In both
calculations the analysis of neutron skin formation is based on
the nuclear form factors, which are well suited for such study
since the diffraction parameters are mainly sensitive to the
nuclear densities in the surface region.

We also show on the example of the neutron-deficient Ni isotopes
the possibility to find a proton skin in a similar way to the
neutron skin. Although the analysis, which was performed in our
paper for this case, uses an alternative criterion to that applied
in \cite{Fuku93}, it indicates a situation close to proton skin
formation in Ni isotopes very close to the proton drip line.
However, the search for the existence of proton skin could be
explored in the most proton-rich nuclei approaching the proton
drip lines of lighter nuclei, where $Z>N$.

In the present work the effects of deformation on the skin
formation are studied in Kr isotopes which are well deformed
nuclei. Taking as an example $^{98}$Kr isotope, we find that the
profiles of the proton and neutron densities, as well as the
spatial extensions change with the direction in both oblate and
prolate shapes. At the same time, the neutron skin thickness
remains almost equal along the different directions perpendicular
to the surface. Same type of calculations have been also performed
on the example of $^{100}$Kr, exhibiting a similar potential
energy curve. In this case, the conclusion concerning neutron skin
thickness on the different directions remains unchanged. We find a
very weak dependence of the neutron skin formation on the
character of deformation. This is a useful
information, worth to be known before complete GCM calculations
are performed because it indicates that no drastic changes in the
neutron skin thickness are expected when such more sophisticated
calculations are performed.

The results obtained in the present paper demonstrate the ability
of our microscopic theoretical method to predict the nuclear skin
in exotic nuclei. They also illustrate the range of the skin sizes
to be expected depending on the adopted skin definition.
More definite conclusions on the emergence of nuclear skin
will be drawn when direct measurements of proton and neutron form
factors, and thus the corresponding proton and neutron densities,
for these nuclei will become available at the upcoming
experimental facilities.

\begin{acknowledgments}
One of the authors (M.K.G.) is grateful for the warm hospitality
given by the CSIC and for support during his stay there from the
State Secretariat of Education and Universities of Spain (N/Ref.
SAB2005--0012). This work was partly supported by the Bulgarian
National Science Fund under Contracts No.~$\Phi$--1416 and
$\Phi$--1501, and by Ministerio de Educaci\'on y Ciencia (Spain)
under Contract No.~FIS2005--00640.
\end{acknowledgments}

\end{document}